\documentstyle[11pt,saltwshop, twoside, psfig, epsf]{article}
\def\edcomment#1{\iffalse\marginpar{\raggedright\sl#1\/}\else\relax\fi}
\reversemarginpar

\begin{document}
\title{Probing Kolmogorov turbulence beyond the Magellanic Clouds: The
 Power of Southern Hemisphere's largest optical telescope (11m), SALT}
 \author{D. L. Block$^1$,
 B.G. Elmegreen$^2$,  D.M. 
Elmegreen$^3$ \& I. Puerari$^4$}
\affil{$^1$ School of Computational and Applied Mathematics, University of the
Witwatersrand,
Johannesburg, 1 Jan Smuts Avenue, Johannesburg 2001, South Africa\\
$^2$ IBM Research Division, TJ Watson Research Center, Box 218, 
Yorktown Heights, NY 10598, USA\\
$^3$ Vassar College, Dept Physics and Astronomy, Poughkeepsie, NY
 12604-0278,
 USA\\
$^4$ Instituto Nacional de Astrof\'\i sica, optica y Electr\'onica,
 Calle Luis Enrique Erro 1, 72840
Tonantzintla, Puebla, M\'exico}

\begin{abstract}
The Hubble classification scheme of galaxies is based on blue-light
appearance. 
Atlases reveal the rich variety of responses of the Population I component
(`the mask') of gas and dust to the underlying, older, stellar population.
However, the Population I component may only constitute 5 percent of the
dynamical mass of the galaxy; furthermore, dusty masks are highly effective
in hiding bars. We firstly discuss the rich duality in spiral
structure, and highlight a near-infrared classification scheme for
spiral galaxies. We next show that images secured with SALTICAM will
be ideally suited to probe key questions such as whether 
the
optical light in the gaseous Population I component is the result of
Kolmogorov turbulence, cascasding from the largest of scales down to
the Nyquist limit. If so, 
the optical
emission in galaxies will be  organized in a global fractal pattern with an
intrinsic 1D power spectrum having a slope of -5/3, or -8/3 in 2D.
\end{abstract}

\section{Introduction}

In Roget's Thesaurus, we find the following:

{\bf Mask:} [noun] screen, cloak, shroud. [verb] to camouflage, to
make opaque, to disguise.

Optically thick dusty domains in galactic disks can completely
camouflage or disguise
underlying stellar structures. {\it Cosmic dust grains act as masks}.
The dust masks obscure whether or not the dust lies in an actual screen
or is well intermixed with the stars. The presence of dust and the
morphology of a
galaxy are inextricably intertwined: indeed, the morphology of a galaxy can
completely change once the Population I disks of galaxies -- the
masks -- are dust penetrated (e.g., Block
and Wainscoat 1991; Block et al.,
1994, 2000).

\begin{figure}
\vspace{12cm}
\caption{The bar in the Large Magellanic Cloud is beautifully portrayed in 
this naked
  eye drawing by Sir John Herschel in 1847. Two dimensional
  power spectra of HI emission in the LMC, spatially spanning
  three orders of magnitude, betrays
  the presence of Kolmogorov
  turbulence (Elmegreen, Kim and Staveley-Smith, 2001), with a power
  law slope of -8/3. SALTICAM will provide unique opportunities to
  examine Kolmogorov turbulence in galaxies beyond the Magellanic Clouds.  
Reproduced from de
  Vaucouleurs 
\& Freeman (1972).}
\end{figure}

The classification of galaxies has traditionally been inferred from
photographs or CCD imaging shortward of the 1$\mu m$ window, where
stellar Population II disks are not yet dust-penetrated. Images
through an $I$ (0.8 $\mu m$) filter can still suffer from
attenuations by dust at the 50\% level. The
NICMOS and
other near-infrared camera arrays offer unparalleled opportunities for
deconvolving the Population I and II morphologies, because the opacity at
$K$ -- be it due to absorption or scattering
-- is always low. The extinction (absorption+scattering) optical depth at $K$
is only 10\% of that in the V-band.

Many years before the advent of large format near-infrared camera
arrays, it
became increasingly obvious from rotation curve analyses that
optical Hubble type is not correlated with the evolved Population II
morphology. This was already evident in the pioneering work of Zwicky
(1957) when he published his famous photographs  
showing the `smooth red arms' in M51.
In the {\it Hubble Atlas} and other atlases showing optical
images of galaxies, we are
looking at masks: at the gas, not the stars,
to which the properties of rotation
curves are inextricably tied.

\section{A duality in spiral structure}

There is a fundamental limit
in predicting what an evolved stellar disk might
look like (Block et al. 2000). The greater the degree of decoupling,
the greater is the uncertainty.
The fact that a spiral might be
flocculent in the optical
is very important, but it is equally important to know whether or not
driving the dynamics is a  
grand design old stellar disk.

Decouplings between stellar and gaseous disks are cited in many
studies including Grosb{\o}l \& Patsis (1998), Elmegreen et al.
(1999),  Block et al. (2000) and Puerari et al. (2000).
The Hubble type
of a galaxy does not dictate its dynamical mass distribution (Burstein
\& Rubin 1985).  This is confirmed by examining  
Fourier spectra, for example, of the evolved disks of NGC
  309 (Sc) and NGC 718 (Sa); these spectra  are almost identical
  (both belong to pitch angle type $\beta$: see Table 1 and Fig. 5 
 in Block and Puerari 1999).

Figure 2 shows how the stellar disks of galaxies can 
{\it
 quantitatively} be classified
in a systematic way.

\begin{figure}
\vspace{12cm}
\caption{Spiral galaxies in the dust penetrated regime are binned
according
to three quantitative criteria: H$m$, where $m$ is the dominant
Fourier harmonic (illustrated here are the two-armed H2 family); 
the pitch angle families $\alpha$, $\beta$ or $\gamma$
and thirdly the bar
strength,
derived from the gravitational torque (not ellipticity) of the
bar. Early type b spirals (NGC 3992, NGC 2543, NGC 7083, NGC 5371
and NGC 1365) are distributed within all three families ($\alpha$,
$\beta$ and $\gamma$). Hubble type and dust penetrated class are 
uncorrelated. One of the enormous potentials of an infrared imager on
SALT will be to examine whether spiral instabilities in stellar disks
are fractal in nature.}
\end{figure}

\section{Power Spectra and Kolmogorov Turbulence}

One of the major processes which structures the interstellar media in
galaxies is mechanical feedback: for example, supersonic winds from
massive stars and multiple supernovae from OB associations. Such
mechanical feedback may generate wind-blown bubbles (e.g. the Rosette:
Block, 1990). Supernovae remnants
as well super-bubbles (in both ionised as well as neutral gas) are two
other well known examples.

The other major process that structures the interstellar medium on
scales of 10 to $\sim$ 500 pc is {\it turbulence}. The standard
reference for turbulent power spectra is the Kolmogorov model,
applicable to homogeneous, isotropic, incompressible and adiabatic 
turbulence.

There are, at present, no good theories about the morphology of
density structures in {\it compressible} turbulence; compressible
turbulence gives a range of structures between incompressible
turbulence and shocks. In a medium that is supersonically turbulent,
the 2D power spectrum slope changes from -8/3 (-2.67) to -3; it is
very 
difficult,
therefore, to distinguish incompressible from shock dominated
turbulence, as the differential signature is not large. Even the most
extreme cloud formation scenarios, where all clouds represent shocks,
would have a power spectrum similar to that of the classic Kolmogorov 
incompressible model.

Self-gravitating clouds of gas have a wide range of masses, from $\sim$
10$^{7} M_{\odot}$ to less than 1$M_{\odot}$. There is no
characteristic or dominant mass for self-gravitating clouds: most
star-forming regions are similar, except for size. It is size which
determines the velocity dispersion of the cloud, as well as its
density,
for a common background 
pressure.

\begin{figure}
\vspace{12cm}
\caption{Forty-eight power spectra are  generated from this HST image
  of the Sb galaxy NGC 4622. The galaxy was analysed in circular
  swaths, from a galactocentric radius of r=10 to r =490 pixels. The
  first circular swath covers (10, 19) pixels, the second one (20, 29)
  pixels, ..., (480, 489) pixels. Adopting a distance to NGC 4622 of 
45.02 Mpc and
  using a HST/WFPC2 scale of 0.09993 arcsec pix$^{-1}$, gives a linear
  scale of 21.81 pc per pixel. The mean radius corresponding to the 
lowermost power spectrum is 15 pixels (or 0.327 kpc); that of the
  uppermost, 485 pixels or 10.577 kpc. Each power spectrum contains
  Fourier modes k=1 to k=1800. We have multiplied each  power spectrum
  by $k^{5/3}$ so that power spectral slopes of -5/3 will appear
  horizontal 
  here. Of particular interest is the -5/3 power law slope 
  within the domain of the probable thickness of its
  galactic disk (for the Milky Way, this is 325 pc). As soon as
  structures are formed with sizes comparable to, or smaller than, the
  disk thickness, the dynamics in the plane and vertically to it are
  no longer independent (see, e.g., Huber and Pfenniger, 2001). 
  HST image
  courtesy R. Buta.}
\end{figure}

The correlated structure
in a turbulent gas is with a power spectrum:

$$ \hat I_c(k)=\sum_{n=1}^{N}\cos(k2\pi n/N)I(n) $$
$$ \hat I_s(k)=\sum_{n=1}^{N}\sin(k2\pi n/N)I(n) $$
$$ P(k)=\hat I_c(k)^2+\hat I_s(k)^2 $$

N is the
number of pixels in the azimuthal scan and k is the wavenumber.
Two dimensional power spectra of the neutral hydrogen (HI) emission of the
Milky Way (e.g. Green, 1993; Dickey et al. 2001) are characterised by power 
laws, with slopes varying from
-2.8 to -3. Elmegreen, Kim
and Staveley-Smith (2001) probed
spatial scales in the Large Magellanic Cloud, 
ranging over three orders of magnitude
(30-4000pc), 
and found a 
2D power law slope to be $\sim$ -2.7 (or equivalently -1.7 in
1D). Velocity and density spectra of the Small Magellanic Cloud are
presented by Stanimirovic and Lazarian (2001);  the spectral
indices (for their 3D power spectra) are -3.3 and -3.4, equivalent to
a 1D power law slope of -1.4.

For classical Kolmogorov turbulence, the implication is that the gas 
is fractal in nature, with the turbulent
kinetic energy cascading to ever smaller scales, according to $E(k)
\alpha k^{-5/3}$. This derives from the assumption of a constant
energy transfer at at scales. 
Values steeper than -5/3 signify progressive energy losses,
such as from compressible or shock dominated turbulence.

If the scale
height in the ambient medium is the typical scale for coherent
star-forming structures, then self-gravity in the gas initiates the
formation of star-forming clouds and star complexes (see Figure 3). 
Why use power spectra instead of conventional 2D Fourier transforms?
For galaxies, the 2D Fourier transform includes structure from the
exponential disk, which is generally not wanted in analyses of
interstellar clouds. Power spectra generated from azimuthal profiles
are, 
however, ideal, since in any given azimuthal swath, there 
will be no systematic radial gradients from the disk.

Elmegreen, Elmegreen and Leitner (2003) have studied some famous 
examples of nearby spirals
(such as NGC 3031=M81), and have generated power spectra of the
optical light in passbands such as B, V and R. Their conclusion is
that the optical light in spiral galaxies (even grand-design ones such
as M81) is the result of
turbulence; the implication therefore, is that young stars follow the
gas as they form. 
They argue that large-scale turbulent motions
may well be generated by sheared gravitational instabilities in the
disk.

SALT will yield the much needed insight into the 
physical scale at which star formation
in galaxies becomes coherent; in our Galaxy, we know this may 
range from several 
hundred parsecs, to a
kpc (Gould's belt, of age $\sim$ 30 Myr, spans $\sim$ 1 kpc).

Its optical imager, SALTICAM, with its default BVRI filters and 
10$\times$10 arcmin field of view, coupled with its high
spatial resolution (15$\mu m$
pixels, 106.7 $\mu m$ arcsec$^{-1}$, $\sim$ 7 pixels per square
second of arc), will be
ideally placed for exploring Kolmogorov turbulence beyond the Magellanic
Clouds, down to the Nyquist limit.

\acknowledgments

We dedicate this paper
to the memory of Bob Stobie, whose vision and dream was SALT.
DLB is indebted to David Buckley for his invitation to participate
in this SALT Workshop. 
This research is supported by the 
Anglo-American Chairman's Fund; a note of deep appreciation is
expressed to the Board of Trustees.

\end{document}